\documentclass{aa} 

\usepackage{graphicx}
\usepackage{txfonts}
\usepackage{epsfig}
\usepackage{ulem}
\usepackage{natbib}

\begin{document}
%
\title{Evidence for CO depletion in the inner regions of gas-rich protoplanetary disks\thanks{Based on observations collected at the European Southern Observatory, Paranal, Chile. (Program ID 079.C-0349A)}}


\author{G. van der Plas\inst{1,2}
  \and M. E. van den Ancker\inst{1}
  \and B. Acke\inst{3}\fnmsep\thanks{Postdoctoral Fellow of the Fund for Scientific Research, Flanders.}
  \and A. Carmona\inst{4,5}
  \and C. Dominik\inst{2,6}
  \and D. Fedele\inst{7}
  \and L.B.F.M. Waters\inst{2,3}
}

\offprints{G. van der Plas, \email{g.vanderplas@uva.nl}}

\institute{European Southern Observatory, Karl-Schwarzschild-Str.2, D 85748 Garching bei M\"unchen, Germany
  \and Astronomical Institute Anton Pannekoek, University of Amsterdam, Science Park 904, 1098 XH Amsterdam, The Netherlands
  \and Instituut voor Sterrenkunde, KU Leuven, Celestijnenlaan 200D, 3001 Leuven, Belgium
  \and ISDC Data Centre for Astrophysics, Ch. d'Ecogia 16, 1290 Versoix, Switzerland
  \and Observatoire de Gen\`eve, Ch. des Maillettes 51, 1290 Sauverny, Switzerland
  \and Institute for Astrophysics, Radbout University, Heyendaalseweg 135, 6525 AJ Nijmegen, The Netherlands
  \and Max-Planck-Institut f\"ur Astronomie, K\"onigstuhl 17, 69117 Heidelberg, Germany 
}

   \date{}

 
  \abstract
   {}
   {We investigate the physical properties and spatial distribution of Carbon Monoxide (CO) gas in the disks around the Herbig Ae/Be stars HD 97048 and HD 100546.}
   {Using high-spectral-resolution  4.588-4.715 $\mu$m spectra containing fundamental CO emission taken with CRIRES on the VLT, we probe the circumstellar gas and model the kinematics of the emission lines. By using spectro-astrometry on the spatially resolved targets, we constrain the physical size of the emitting regions in the disks.}
   {We resolve, spectrally and spatially, the emission of the $^{13}$CO v(1-0) vibrational band and the $^{12}$CO $v=1-0, v=2-1, v=3-2$ and $v=4-3$ vibrational bands in both targets, as well as the $^{12}$CO $v=5-4$ band in HD 100546. Modeling of the CO emission with a homogeneous disk in Keplerian motion, yields a best fit with an inner and outer radius of the CO emitting region of 11 and $\geq$ 100 AU for HD 97048. HD 100546 is not fit well with our model, but we derive a lower limit on the inner radius of 8 AU. The fact that gaseous [OI] emission was previously detected in both targets at significantly smaller radii suggests that CO may be effectively destroyed at small radii in the surface layers of these disks}
   {}


   \maketitle

\section{Introduction}
We present a study of warm carbon monoxide (CO) gas in two circumstellar disks surrounding young intermediate-mass stars. CO is commonly detected in disks \citep{2003ApJ...589..931N}, and often used as a tracer of the physical conditions in the disk and the kinematics of the gas. Ro-vibrational transitions of warm (T = 120-1010 K) CO gas have already been found by \citet{1990ApJ...363..554M} in 8 out of 9 surveyed embedded infrared sources. \citet{2007ApJ...659..685B} had a 100 \% detection rate for the ro-vibrational CO transitions in the disks around 9 intermediate mass stars with optically thick inner disks, while their detection rate is only 1 out of 5 for disks with an optically thin inner disk, a case they explain by UV fluorescence. \citet{2004ApJ...606L..73B} have detected CO gas in 5/5 Herbig Ae stars and recently, \citet{2008ApJ...684.1323P} have spatially and spectrally resolved the 4.7 $\mu$m CO ro vibrational lines in the disks around 3 young solar mass stars with known dust gaps or inner holes, and detected CO well inside the dust gaps in all disks. 

HD 97048 (Sp. type A0pshe, \citep{1998A&A...330..145V}) is a well-studied nearby Herbig Ae star, whose spectral energy distribution (SED) displays a large infrared excess above the emission from the star, commonly explained by emission from warm dust in a circumstellar disk. The disk is a flaring disk according to the classification proposed by \citet{2001A&A...365..476M}. \citet{2006Sci...314..621L} resolved the disk around HD 97048 in the mid-IR and determined the flaring angle. PAH emission from radii up to 200-300 AU has been detected by \citet{2004A&A...418..177V}. \citet{2004ApJ...614L.129H} have resolved emission from the 3.43 $\mu$m and 3.54 $\mu$m diamond bands from within the inner 15 AU in the disk. Circumstellar gas has also been detected, e.g. by \citet{2006A&A...449..267A} who detect spatially and spectrally resolved [OI] emission from a few tenths to several tens of AU. HD 97048 is one of the only two Herbig stars where the $H_2$ S(1)  line at 17.035 $\mu$m  has been detected \citep{2007ApJ...666L.117M}, coming from warm (T$_\mathrm{ex} ~ \leq ~ 570$ K) gas within 35 AU.

HD 100546 (Sp. type B9Vne \citep{1998A&A...330..145V}) also has a SED characteristic for a flared disk. Its IR spectrum indicates a high crystalline silicate dust fraction in the disk surface layers  \citep{2003A&A...401..577B}.  There is observational evidence for a disk gap, or 'second wall', at $\approx$ 10 AU from the central star.  This was derived based on the SED \citep{2003A&A...401..577B}, and confirmed by nulling interferometry \citep{2007ApJ...658.1164L}. The [OI] 6300 \AA~  line was studied by \citet{2006A&A...449..267A}, who also find evidence for a disk gap at that distance,  possibly induced by a very low mass stellar companion or planet of  $\approx ~ 20 ~ M_\mathrm{jup}$, or a planetary system. The latter authors spatially and spectrally resolve the [OI] emission, and determine it to originate from a few tenths to several tens of AU. \citet{2007A&A...476..279G} have detected PAH emission from 12 $\pm$ 3 AU, and the circumstellar material has been spatially resolved many times, revealing rich structures in the disk. 

In this paper, we present high spectral resolution observations with milli-arcsecond spatial precision of CO emission originating from the disks around HD 97048 and HD 100546. 
\section{Observations and data reduction}

\begin{table}
\small
\begin{minipage}[t]{\columnwidth}
\caption{Astrophysical parameters of the programme stars}
\label{table:stellar_parameters}
\centering
\renewcommand{\footnoterule}{}  
\begin{tabular}{llccccccc}
\hline \hline

HD & $\log T_{\rm eff}$ & log $L_{\rm bol}$  & M & Dist. & i & PA  \\    
& log [K] & log [$L_{\odot}$]  & [$M_{\sun}$] & [pc] & [$^{\circ}$] & [$^{\circ}$] \\

\hline

97048  & 4.00  & 1.42 & 2.5$^{\pm 0.2}$ & 180\footnote{\citet{1998A&A...330..145V} $^b$\citet{2006Sci...314..621L} $^c$\citet{2006A&A...449..267A} $^d$\citet{2007ApJ...665..512A}}& 42.8$^{+0.8, {\it b}}_{-2.5}$ & 160$^{\pm 19, {\it c}}$\\ 
100546 & 4.02 & 1.62 & 2.4$^{\pm 0.1}$ & 103$^{{\it a}}$ & 42$^{\pm 5, {\it d}}$ & 145 $^{\pm 5, {\it d}}$\\ 

\hline
\end{tabular}
\end{minipage}
\end{table}

High-spectral-resolution (R $\approx$ 94000, determined from telluric lines) spectra of HD 97048 and HD 100546 were obtained on the early morning of June 16$^{th}$ 2007 with the VLT cryogenic high-resolution infrared echelle spectrograph (CRIRES\footnote{http://www.eso.org/sci/facilities/paranal/instruments/crires/}, \citet{2004SPIE.5492.1218K}).  
Adaptive Optics were used to optimize the signal-to-noise and the spatial resolution of the observations. These cover the wavelength range of 4.588 - 4.711  $\mu$m, and contain many fundamental ro-vibrational transitions of the CO molecule. The observations were made with a slit width of 0.2", and with the slit rotated along the parallactic angle. HD 97048 and HD 100546 were observed with 2 wavelength settings, each for 10 and 3 minutes respectively, and the slit Position Angle (PA) during the observations spans a range of 41.1 - 50.3 degrees for HD 97048 and 74.8 - 78.2 degrees for HD 100546.  This spread in PA, close to the minor axis of the disks, is not optimal for our astrometry purpose, but was enforced by the observational circumstances. The data were reduced using the CRIRES pipeline V1.5.0\footnote{http://www.eso.org/sci/data-processing/software/pipelines/index.html}, which performs wavelength calibration, background subtraction and flatfield correction.

\begin{table}
\setlength{\tabcolsep}{1.1mm}
\caption{Log of spectroscopic observations at June $16^{th}$ 2007}              
\label{table:dataset}      
\centering                                      
\begin{tabular}{ c c c c c c}          
\hline\hline                        
Object & UT start & Ref. wl. & N exp & Exp. time & PA range\\
& hh:mm:ss & [nm] & & [s] & [$^{\circ}$]\\    
\hline   			     
  HD 97048   & 00:41:15 & 4662.1 & 20 & 15 & 41.1 - 45.6\\
             & 00:58:43 & 4676.1 & 20 & 15 & 46.0 - 50.3\\
\hline   			     
  HIP 052419 & 02:36:14 & 4662.1 & 2 & 15 & 85.1 - 85.3\\
             & 02:38:47 & 4676.1 & 2 & 15 & 85.7 - 85.9\\
\hline                                    
  HD 100546  & 02:58:10 & 4662.1 & 6 & 15 & 74.8 - 76.3\\    
             & 03:05:44 & 4676.1 & 6 & 15 & 76.7 - 78.2\\  
%
\hline                                    
  HIP 061585 & 03:49:57 & 4662.1 & 2 & 15 & 72.5 - 72.7\\    
             & 03:52:30 & 4676.1 & 2 & 15 & 73.1 - 73.3\\    
\hline                                             
\end{tabular}
\end{table}

\begin{figure*}
  \centering
\includegraphics[width=19cm,height=7cm]{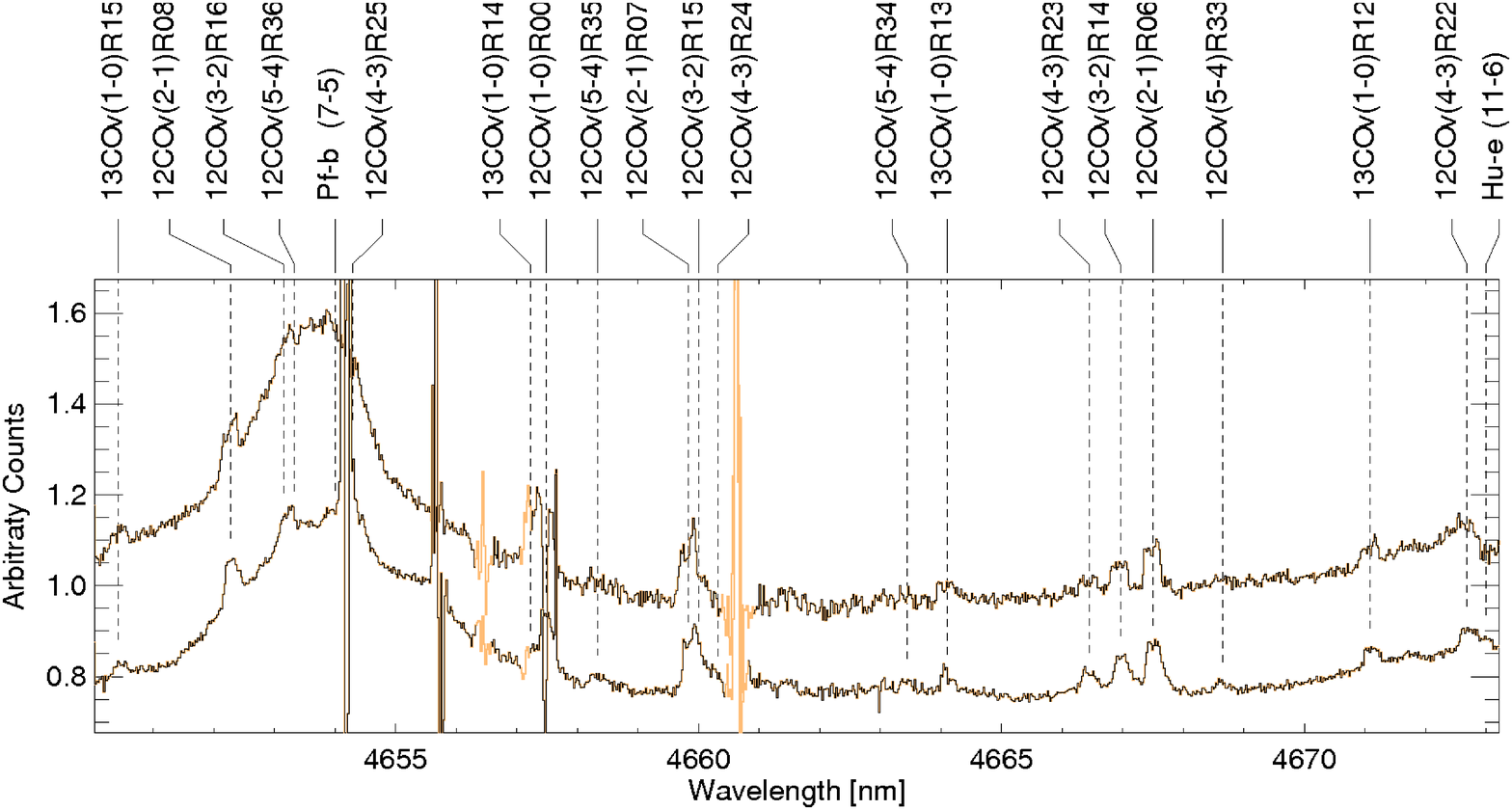}

  \caption{Reduced CRIRES spectra of HD 97048 (top) and HD 100546 (bottom). The black spectrum has been used for the line measurements, whereas the red line shows the areas that have been ignored due to large errors, mostly caused by the correction for saturated telluric lines. The line identifications plotted above show detections of the isotope ($^{12}$CO and $^{13}$CO), the vibrational transition (e.g. $v=1-0$), and the rotational transition ($R(xx)$ or $P(xx)$). The broad hydrogen recombination lines Pfund $\beta$ at 4654 nm and Humphreys $\epsilon$ at 4673 nm are also shown.}
     \label{fig:spec_HD97048}%
\end{figure*}

To correct for telluric absorption we have observed a telluric standard star directly after each science target: HIP 052419 (Sp. type B0V) for HD 97048 and HIP 061585 (Sp. type B2IV) for HD 100546. Each standard was chosen to be as close as possible on the sky so that their spectra are affected by similar atmospheric conditions. We compared their spectra to an appropriate \citet{1991ppag.proc...27K} stellar atmosphere model to determine the instrumental response. The optical depth of the telluric lines of the standard was scaled to that of the science target. The science spectra were derived by division with the instrumental response. A sub-pixel wavelength shift was manually added to reduce telluric residuals (i.e. spikes due to minor wavelength mismatches). Some telluric absorption lines are fully saturated, causing problems with the division of the spectra (i.e. division by 0). In the further data reduction, we will disregard these areas. We have shifted the spectrum to center the $v=2-1$ lines on their rest wavelength. The spectra are shown in Fig. \ref{fig:spec_HD97048}.

\section{Observational Analysis}
\label{sec:analysis}

\begin{figure}
   \centering                  
   \includegraphics[width=9cm]{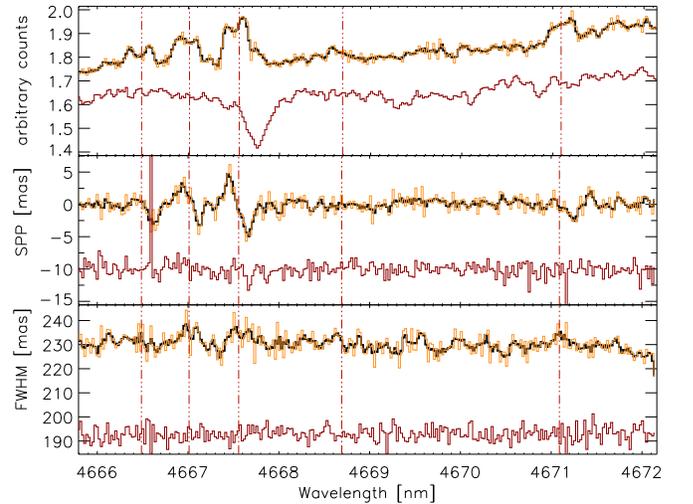}
   \caption{In all three windows the thick black line represents the spectrum of HD 97048 (orange thin line) smoothed over 2 pixels, and the red line denotes the same data for the telluric standard. Shown with the vertical dotted lines are the locations of the (from left to right) $^{12}$CO $v=4-3 R(23)$, $v=3-2 R(14)$, $v=2-1 R(06)$, $v=5-4 R(33)$ and $^{13}$CO $v=1-0 R(12)$ emission lines. {\it Top:} Close up of the spectrum of HD 97048 (as shown in Fig. \ref{fig:spec_HD97048}). {\it Middle:} The SPP of HD 97048 and of the telluric standard, shifted down by 10 mas. {\it Bottom:} The spatial FWHM of the PSF for each spectral row for HD 97048 and the telluric standard.}
         \label{fig:HD97048_resolved}
   \end{figure}

\begin{figure}
   \centering
   \includegraphics[width=9cm]{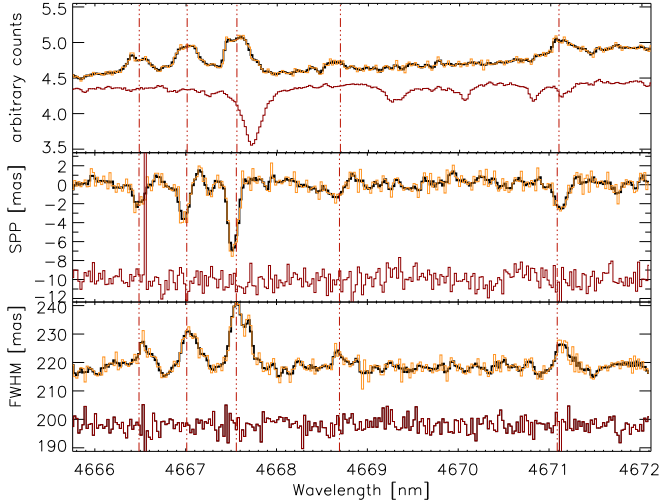}
   \caption{Same as figure \ref{fig:HD97048_resolved} for HD 100546}
         \label{fig:HD100546_resolved}
   \end{figure}

To check whether the targets are spatially resolved, we perform spectro-astrometry on the reduced 2D frames. We determine the location of the photocenter of the PSF and the spatial FWHM as function of wavelength (or equivalently velocity for each line), by fitting a Gaussian to the spatial profile in each spectral column. The relative position of the photocenter to the continuum, the Spatial Peak Position (SPP), is a measure for the spatial offset of the emitting region compared to the continuum. Our observations are made with the slit positioned over a small range in position angles. We note that \citet{2008ApJ...684.1323P} have observed three nearby T Tauri stars at 6 different PA's with CRIRES and the same observational set-up as employed by us, and find, after pairwise subtracting parallel and anti parallel spectro-astrometric signals, that no telluric feature produces false spectro-astrometric signals in excess of 0.2-0.5 mas. Since we do not possess anti-parallel spectra, we show the telluric absorption in Figs. \ref{fig:HD97048_resolved} and \ref{fig:HD100546_resolved}, and note, by comparing with CO lines un-affected by telluric absorption, that their effect on the astrometric signal is small. 
We compare the SPP in the CO lines to the SPP of the underlying continuum (SPP=0 by definition). If the target is resolved, and the photocenter of the CO emission region is offset to that of the dust continuum emission from the disk, the SPP is expected to be non-zero. The photocenter displacement can be derived from the measured SPP if we correct for dilution by the continuum flux, by multiplying the SPP with a factor $1 + F_c(v)/F_l(v)$, with $F_c(v)$ and $F_l(v)$ the continuum and line flux at velocity $v$ \citep{2008ApJ...684.1323P}.  

The Gaussian fit to the PSF also yields a spatial FWHM. If the continuum is unresolved, this quantity is a measure for the achieved effective spatial resolution in the direction parallel to the slit. When resolved, the FWHM allows us to determine the spatial extent of the emission. Outside the spectral lines it then is a direct probe for the size of the continuum emitting disk, whilst inside the lines it is a function of both the line- and continuum emission, and its value a lower limit to the CO emission. Thus, measuring the SPP and the FWHM of the PSF in the continuum and the CO lines allows us to determine [1] the size of the continuum emitting region, and [2] the spatial offset of each CO line with respect to the continuum. This procedure is shown in Figs. \ref{fig:HD97048_resolved} and \ref{fig:HD100546_resolved}. 

\section{Results}
\label{sec:results}



We detect in the spectrum of HD 97048 the $^{13}$CO $v=1-0$, as well as in the $^{12}$CO $v=1-0$ to $v=4-3$ vibrational bands. All lines are spectrally resolved with a remarkably constant FWHM of 15 $km ~ s^{-1}$.  The Half Width at Zero Intensity (HWZI) is  more difficult to estimate due to the continuum S/N, but we make a conservative estimate of 15 $km ~ s^{-1}$ corresponding to an inner radius of 4.5 AU, assuming the stellar parameters and inclination as given in Table \ref{table:stellar_parameters}. Average line profiles and astrometric signals of the $^{12}$CO $v=1-0$ to $v=3-2$ are shown in Fig. \ref{fig:av_line_hd97048_all} and Table \ref{table:spatial transitions}. 
 The 4.6 $\mu$m continuum emission of the disk around HD 97048 is resolved. Its FWHM of 0.231" $\pm$ 0.004" corresponds to 23 $\pm$ 3 AU, after correcting for the FWHM of the PSF. The latter  (0.192" $\pm$ 0.001")  was estimated from 10 unresolved telluric standards, all sufficiently bright for the AO loop to close at maximum frequency, observed with the same instrumental settings, and within 2 days of our observations. The red- and blue- shifted wings of the SPP form a sinusoidal signal, typical for an inclined circumstellar disk. 

In the spectrum of HD 100546 we detect the $^{13}$CO $v=1-0$, as well as in the $^{12}$CO $v=1-0$ to $v=5-4$ vibrational bands. All lines are spectrally resolved and have a FWHM of 15 $km ~ s^{-1}$, irrespective of the transition.  A conservative estimate for the HWZI is 17.5 $km ~ s^{-1}$ corresponding to an inner radius of 3.3 AU. The average line profiles and astrometric signals  of the  $^{12}$CO $v=1-0$ to $v=4-3$ transitions are shown in Fig. \ref{fig:av_line_hd100546_all} and their values in Table \ref{table:spatial transitions}. 
 The 4.6 $\mu$m continuum emission of the disk around HD 100546 is resolved. Its FWHM of 0.218" $\pm$ 0.002"  corresponds to 11 $\pm$ 1 AU, after correcting for the FWHM of the PSF. The SPP is very asymmetric, i.e the negative shift of the blue wing in the SPP is larger than the positive SPP of the red wing. This asymmetry is also seen in the FWHM  and is more pronounced for the lower vibrational transitions. 

\begin{table}
\small
\begin{minipage}[t]{\columnwidth}
\caption{Spatial information for the {\it averaged} vibrational transitions shown in Figs \ref{fig:av_line_hd97048_all} and \ref{fig:av_line_hd100546_all}, and the radii used to model the kinematics.}
\label{table:spatial transitions}
\centering
\renewcommand{\footnoterule}{}  
\begin{tabular}{lllll|l}
\hline \hline
HD 97048    & cont.  & v(2-1)     & v(3-2)   & v(4-3) & $R_{\rm in, out}^{\rm model}$  \\ 
              
\hline
SPP [AU]  & 0       &  14.8$^{\pm 0.9}$ & 16.5$^{\pm 2.7}$    & - & 11   \\
FWHM [AU]\footnote[1]{Note that this FWHM is a function of continuum + line emission, but that the SPP has been corrected for the continuum dilution, and thus is a direct function of the CO emission. All values are directly derived from the 2D spectrum, and not the maximum value resulting from an alignment of the slit with the semi-major axis.} & 21.2$^{\pm 1.6}$ &  $\geq$23.8$^{\pm 1.1}$ & $\geq$23.0$^{\pm 1.1}$ & - & 100  \\

\hline
HD 100546    &   &      &    &  &   \\ 
\hline
SPP [AU]  & 0     &  9.6$^{\pm 0.5}$ & 9.4$^{\pm 0.5}$    & 9.9$^{\pm 1.5}$   & 8  \\
FWHM [AU] & 10.0$^{\pm 0.9}$ & $\geq$13.7$^{\pm 0.6}$ & $\geq$13.1$^{\pm 0.6}$  & $\geq$11.5$^{\pm 0.7}$   &  100 \\

\hline
\end{tabular}
\end{minipage}
\end{table}

%


\section{A Simple Model}
\label{sec:model}

\begin{figure}
   \centering
   \includegraphics[width=9cm]{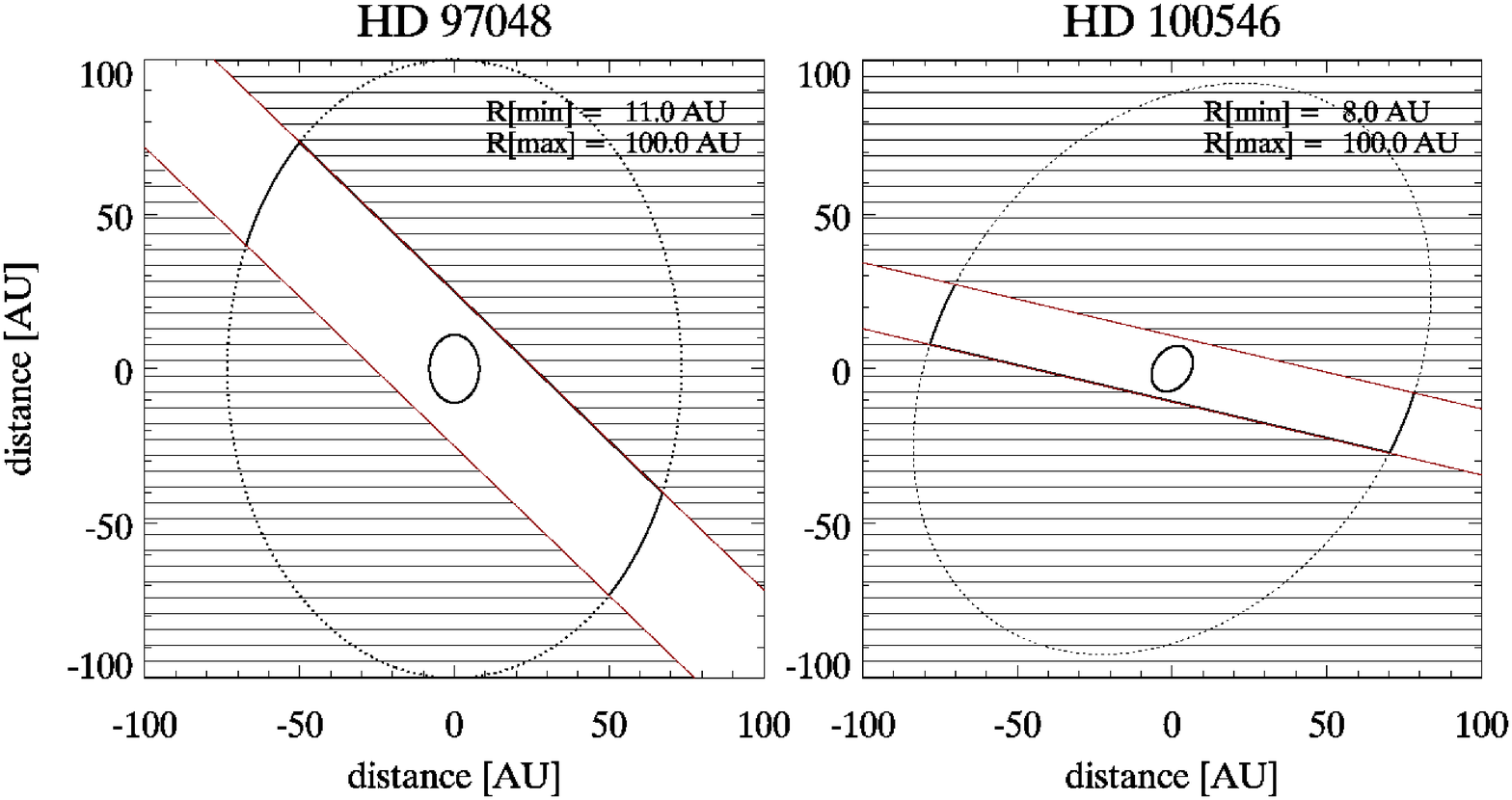}
   \caption{Projected images of the simulated CO with the best fit parameters for HD 97048 and HD 100546, together with the slit of the spectrograph at the time of observation. North is up, East is left.}
         \label{fig:diskimage}
   \end{figure}

\begin{figure}
   \centering
   \includegraphics[width=9cm]{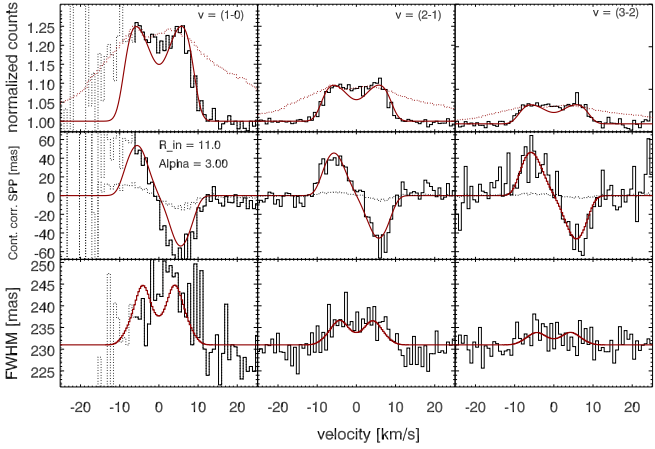}

   \caption{{\it Top} The average line profiles of the HD 97048 $v=1-0, v=2-1$ and $v=3-2$ transitions, created with  5 ($5 < \mathrm{J} < 7$), 12 ($5 < \mathrm{J} < 14$) and 7 ($14 < \mathrm{J} < 21$) lines respectively, are shown with the black histogram. The blue wing of the  $v=1-0$ transition is contaminated by telluric CO lines, shown with dots. Over plotted is the 6300 \AA~ [OI] emission (red dotted line), and the model fit to the CO emission with the solid red line. {\it Middle} The corresponding, continuum corrected (section \ref{sec:analysis}, the uncorrected SPP is shown with the dotted line), average astrometric signal. The model is over plotted in red. {\it Bottom} The average spatial FWHM in black, with the model over plotted in red.}
         \label{fig:av_line_hd97048_all}
   \end{figure}

A powerful method to determine the extent of the emitting gas is to model the kinematics of the line- and astrometric profiles. We construct the average profiles for the $v=1-0$ to $v=4-3$  vibrational bands in Figs. \ref{fig:av_line_hd97048_all} and \ref{fig:av_line_hd100546_all},  and fit these to the model  shown in Fig. \ref{fig:diskimage}. In this model, the gas is in Keplerian orbit in a flat disk with known inclination and PA around a star with known stellar mass and distance. Further, the intensity of the emission decreases as $I(R) = I_\mathrm{in}\left(\frac{R}{R_\mathrm{in}}\right)^{-\alpha}$, with $I_\mathrm{in}$ the intensity at the inner radius $R_\mathrm{in}$, and $R$ the radial distance from the star \citep{2007A&A...476..853C}. The astrometric signal is consequently determined by fitting a Gaussian to the simulated PSF. Because the stellar parameters are all well constrained, the free parameters are $\alpha$, $R_\mathrm{in}$, and $R_\mathrm{out}$.  $\alpha$ describes the change in intensity of the CO emission as function of radius, where $\alpha$ = 2 reflects an intensity directly proportional to the stellar radiation field assuming the surface density has no radial dependence. We compare the line profile, the SPP and the FWHM with their simulated counterparts for the parameter-space $ 1 \mathrm{ AU} \leq R_\mathrm{in} \leq 29 \mathrm{ AU} $,  $ 25 \mathrm{ AU} \leq R_\mathrm{out} \leq 200 \mathrm{ AU} $ and $0 \leq \alpha \leq 5.5$, and quantify the best fit by minimizing the reduced $\chi^2$. We estimate the variance on the data from the continuum adjacent to the line, and determine the best fit from the $v=2-1$ line profile, because the $v=1-0$ data is affected by telluric absorption, and because this transitions has a higher S/N compared to the higher vibrational transitions. Since the continuum corrected SPP is similar for each vibrational transition within the separate stars, we assume that all vibrational transitions have the same spatial origin. For HD 100546, and in lesser extent HD 97048, our model produces a more distinctly double peaked line profile than the observed profiles. This phenomenon is also observed for 9 out of 60 protoplanetary disks (Bast et al., in prep.), and suggests there is an extra low projected velocity gas component present in those disks. We have repeated the fitting procedure for only the part of the line profile sensitive to the inner part of the disk, the line wings ($\left|V_\mathrm{proj}\right| \geq 4.5 km ~ s^{-1}$), and find no difference for the best fit.%



\begin{figure}
   \centering
   \includegraphics[width=9cm]{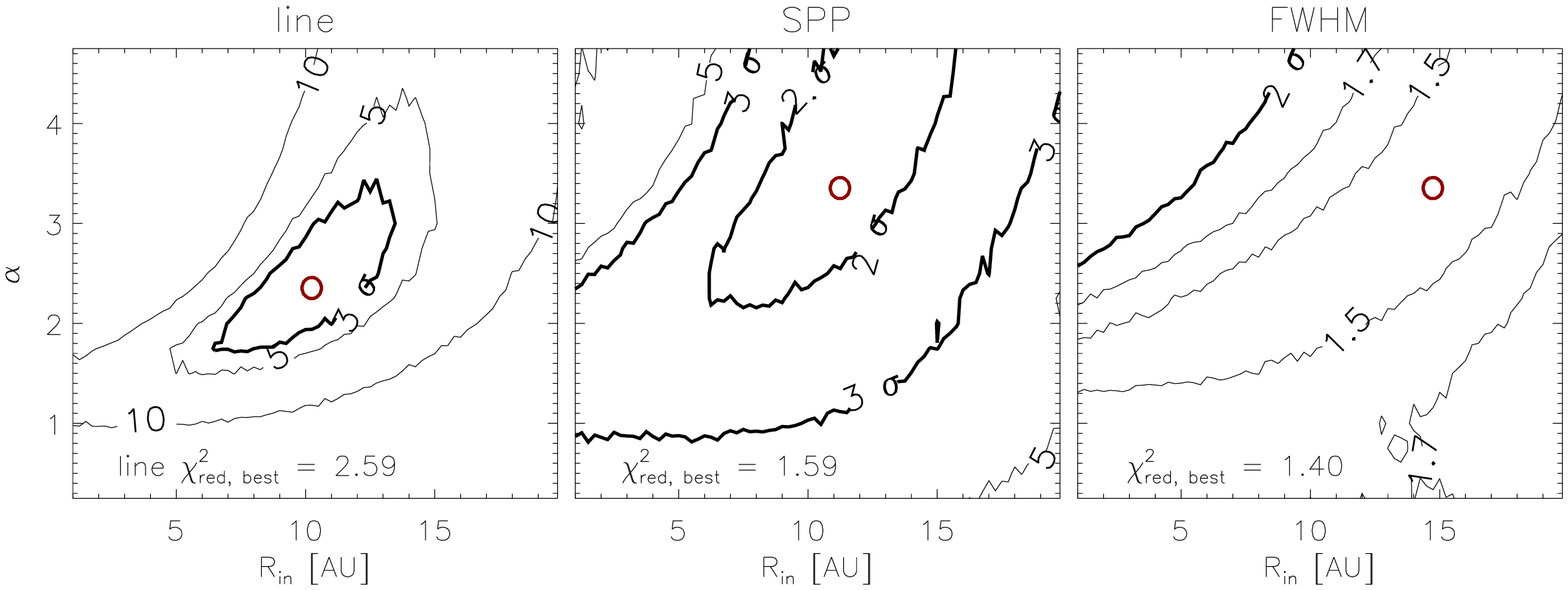}
   \caption{$\chi^2_\mathrm{red}$ surfaces for the best fitting disk model for, from left to right, the line profile, SPP and FWHM of the $v=2-1$ transition of HD 97048. The minimal $\chi^2_\mathrm{red}$ for each fit is given in the panels, and the 1, 2 and 3 $\sigma$ lines represent  $\geq$68$\%$, $\geq$90$\%$ and $\geq$99$\%$ confidence levels. The best fit parameters for each panel are shown by a red circle. To better show the $\chi^2_\mathrm{red}$ surface landscape we have overplotted the $\chi^2_\mathrm{red}$ = 5, 10 (left two panels, $\chi^2_\mathrm{red}$ = 1.5, 1.7 in the right panel) contours.}
         \label{fig:hd97048_chisq}
   \end{figure}

\begin{figure}
   \centering
   \includegraphics[width=9cm]{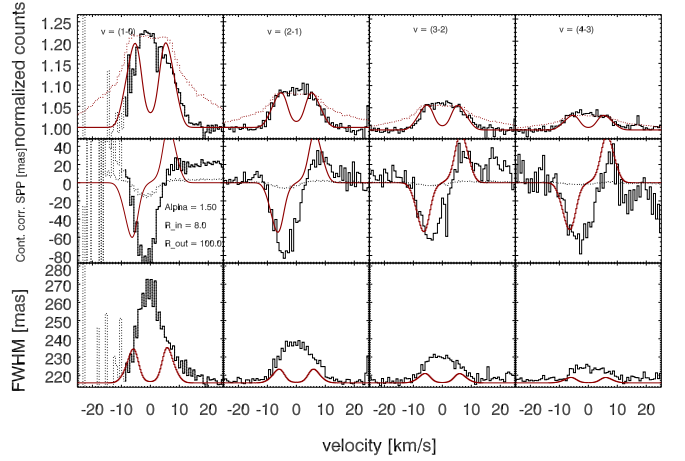}
   \caption{Same as Fig. \ref{fig:av_line_hd97048_all} for HD 100546, but the $v=1-0, v=2-1, v=3-2$ and $v=4-3$ profiles are made from 11 ($0 < \mathrm{J} < 7$), 12 ($5 < \mathrm{J} < 14$), 9 ($11 < \mathrm{J} < 26$) and 12 ($19 < \mathrm{J} < 34$) lines respectively.}
         \label{fig:av_line_hd100546_all}
   \end{figure}

The best fit to the $v=1-0$, $v=2-1$, and $v=3-2$ vibrational transitions of HD 97048, are $R_\mathrm{in}$ $=$ 11 AU, $R_\mathrm{out}$ $\geq$ 100 AU and $\alpha$ $=$ 3.00. We show the $\chi^2_\mathrm{red}$ surfaces for the $v=2-1$ fit to the line profile, the SPP and the FWHM in Fig. \ref{fig:hd97048_chisq}. These values have been derived using a PA of 180$^\circ$, estimated from direct imaging of the circumstellar disk (Fig. S1 of \citet{2006Sci...314..621L}), compared to 160$^{\pm 19 \circ}$, found by \citet{2006A&A...449..267A} using astrometry. Using the latter value makes the simulated line profile slightly more double peaked, and thus increases the minimal $\chi^2_\mathrm{red}$, but does not change the best fit parameters.  

The spatial distribution of gas in HD 100546 is more complex than the 'simple' homogeneous disk in Keplerian motion we model, and the best fit values for each transition vary, with a lowest value of $R_\mathrm{in}$ $\geq$ 4 AU and $\alpha$ $\geq$ 1.5. If we assume that all vibrational states have the same spatial distribution and origin, and we only fit the line wings, a more likely value of  $R_\mathrm{in}$ $=$ 8 AU and $\alpha$ $=$ 1.5 is found. The poor quality of the fit is possibly due to complex disk geometry and gas kinematics, related to the presence of a  disk gap and consequent ``second wall''  at 10 AU \citep{2003A&A...401..577B}. The average profiles of all vibrational bands have a FWHM of 15 $km ~ s^{-1}$, hinting that they share a common spatial origin. The [OI] emission has much wider wings (FWHM = 24 and 25 $km ~ s^{-1}$ for HD 97048 and HD 100546), corresponding to a smaller $R_\mathrm{in}$ (80 \% of the [OI] emission emanates from the region between 0.8 and 20 AU, Fig. 7 in \citet{2006A&A...449..267A}).


\section{Discussion}




Fundamental CO emission from Herbig stars, and gaps therein, have been detected in more circumstellar disks, e.g. in SR 21 with an  $R_\mathrm{in}$ of 7.6 $\pm$ 0.4 AU for the  $^{12}$CO v(1-0) line \citep{2008ApJ...684.1323P}, while there is an inner dust component at $\leq$ 0.5 AU, followed by a dust gap up to 18 AU \citep{2007ApJ...664L.107B}. \cite{2006ApJ...652..758G} find an inner clearing in the CO emission at 11 $\pm$ 2 AU around HD 141569, and suggest the cavity most likely is cleared by photo-evaporation working together with viscous accretion. \cite{2007ApJ...659..685B} observe the same star, find CO emission starting between 6 and 15 AU, and explain the emission as due to UV fluorescence. \cite{2004ApJ...606L..73B} study a sample of 5 Herbig stars, not known to have disk gaps or inner holes, and find that the CO gas traces both collisional driven emission in the inner ($<$ 0.5 - 1 AU) disk, as well as CO gas excited by resonance fluorescence (R $<$ 50-100 AU). These observations suggest that CO depletion in the inner disk is associated with large disk gaps or inner holes. Finally, in all cases discussed above (except SR 21 for which no data exists), the [OI] emission, when detected, arises at similar or higher velocities than the CO fundamental ro-vibrational emission. 

Contrary to the above, the lack of CO emission at small radii in HD 97048 and HD 100546 can not be caused by a gas-free inner disk. [OI] 6300 \AA~ emission, a tracer of  OH molecular gas \citep{2005A&A...436..209A} in the surface layers of flared disks, and induced by the stellar UV field, has been observed for both stars by \citet{2006A&A...449..267A}, with  $80 \%$ of the emission  originating between  $0.8 < R < 20$ AU.  The SEDs of our targets also display large NIR excesses, typical for hot dust close to the star \citep{2004A&A...426..151A, 2003A&A...401..577B}. In the cases of HD 97048 and HD 100546, the absence of CO emission at Small radii can thus not be attributed to the paucity of matter in the inner disk, nor to the absence of UV flux. Disk holes and/or large gaps therefore are not the only cause of CO depletion in the inner disk. Here, we suggest that CO is efficiently destroyed in the inner disk. A possible mechanism is the photo dissociation of CO and the subsequent formation of simple organic species such as C$_2$H$_2$, HCN and CH$_4$ \citep{2008A&A...483..831A}. Further analysis (e.g. observations of hot transitions of carbon-bearing molecules) is needed to determine the reason why the CO gas is not detected in the inner regions, and may offer us a glimpse of the molecular chemistry of these inner disks.

The similar line widths and shapes for all vibrational states in HD 97048, as well as the similar continuum corrected SPP, suggest a common excitation mechanism and location. The best fit parameter for the radial dependency of the CO emission intensity $\alpha$ = 3.0  can guide models that aim to explain the highly excited vibrational line emission, such as UV fluorescence. The similar line shape, continuum corrected SPP and FWHM of the higher ($v=3-2$ and $v=4-3$) vibrational transitions in HD 100546 also suggest a common excitation mechanism and location, but the $v=2-1$ and especially $v=1-0$ vibrational transitions deviate significantly. The asymmetry in the data on HD 100546, and the poor fit to the model, point to a more complicated disk geometry, in agreement with finding of \citet{2003A&A...401..577B} and \citet{2006A&A...449..267A} 


\acknowledgements{The authors of this paper wish to thank the Paranal nighttime astronomer E. Valenti for her cheerful assistance making the observations, and also the anonymous referee for helping improving this paper.}

\end{document}